\documentclass[aps,prl,preprint,groupedaddress,amsmath,amssymb,showpacs]{revtex4}
\usepackage{graphicx}
\usepackage{amsmath}

\begin{document}

\title{Quasi-paraxial theory for coupled unstable cavities I: formal development}


\author{A. Aiello}
\author{J. P. Woerdman}
\affiliation{Huygens Laboratory, Leiden University, P.O. Box 9504,
Leiden, The Netherlands}
\date{\today}
\begin{abstract}
We present a formal wave theory for the calculation of the
spectrum and the eigenmodes for a certain class of  ray-chaotic
optical cavities introduced by A. Aiello, M. P. van Exter, and J.
P. Woerdman [quant-ph/0307119].
\end{abstract}
%
%
\maketitle
%
%
In a previous paper 
 \cite{Aiello03a},
 we presented  a theoretical  model for a composite optical cavity made of standard laser
mirrors; the cavity consists  of a suitable combination of stable
and unstable cavities as shown in Fig. 1. By using numerical
simulation we were able to demonstrate that such a cavity displays
classical (ray) chaos, which may be either soft or hard, depending
on the cavity configuration. In this paper we want to go a step
further by addressing the behavior of the chaotic cavity in a wave
regime (or, loosely speaking, in a ``quantum'' regime
\cite{StockmannBook}). More precisely, in this paper we present a
{\em formal} theory for two coupled unstable cavities. We show
that it is possible to introduce an {\em unitary} coupling which
accounts both for direct transmission and diffraction (which
occurs from the edges of the convex mirrors in our cavity) by
using a suitable scattering operator (see Eqs.
(\ref{eq:4}-\ref{eq:8}) below).

A standard  two-mirror {\em stable} resonator is a {\em
geometrically open} system but because of its stability it is {\em
closed} both from ray \cite{Hersch00a} and wave point of view. In
other words, a typical gaussian-beam-like mode in such a resonator
is confined both longitudinally (that is along the axis of the
resonator) and transversally (that is along the two directions
orthogonal to the axis) by the focussing action of the two
mirrors. Because of this confinement a stable resonator has a
discrete spectrum; in paraxial approximation this spectrum can be
classified in a ``longitudinal'' part which depends only on the
length of the cavity and in a ``transversal'' one which depends
also from the radii of curvature of the two mirrors. Here we are
interested mainly in the transversal part.
\\
Efficient  methods to calculate the spectrum and the eigenmodes of
hard-edged unstable cavities were developed in the last 30 years;
particularly notable is the asymptotic theory created by Horwitz
\cite{Horwitz73a} and Southwell \cite{Southwell86a}. However, in
spite of this long hystory, surprising properties of these
eigenmodes were discovered recently \cite{Karman98a,
Karman99a,Berry01a,Berry01b}. For instance, the Horowitz-Southwell
theory has been exploited and slightly modified by Berry {\em et.
al.} to investigate both the fractal nature of the cavity
eigenmodes \cite{Berry01b} and the occurrence of the Petermann
excess-noise factor \cite{Berry03a}.
In this paper we apply Berry's theory to our composite cavity,
thus generalizing some of the results presented in
\cite{Berry01b}. From a mathematical point of view, the main
difference between the theory for a conventional unstable cavity
and our composite system, is that in the former case the operator
which accounts for the modes propagation inside the unstable
cavity is not unitary because of the losses from the edges of the
smallest mirror. As we shall show later, in our case the two
round-trip operators describing the mode propagation in the two
half cavities shown in Fig. 1 remain non-unitary but the operator
describing the motion in the overall cavity is unitary because the
whole cavity is stable ($L < 2R$).

In this paper we restrict our attention to two-dimensional
cavities with one-dimensional mirrors ({\em strip} resonators).
Following Berry \cite{Berry03a} it is convenient to introduce from
the beginning a ``quantum-like'' vector-space notation writing the
modes of the field as kets in a linear space defined by the
propagation operator $\hat{K}$ whose coordinate representation is
given by the Huygens' integral in the Fresnel approximation
\cite{SiegmanBook}. Within this formalism, the transversal mode
profile $u(y)$ calculated in an arbitrary plane $z =$ const. can
be considered as the coordinate representation of a field state
$|u\rangle$ depending on the longitudinal coordinate $z$ which is
considered as a parameter (exactly as the time in the
Scr\"{o}dinger equation):
\begin{equation}\label{eq:0}
  \langle y | u \rangle \equiv u(y).
\end{equation}
In order to describe the dynamics of each sub-cavity and the
coupling between them, we introduce a set of four fields $u_1,
u_2$ and $v_1, v_2$ defined in the reference plane $z = 0$
following the scheme illustrated in Fig. 2. Then the propagation
in the left and right side of the whole cavity can be described by
introducing the operators $\hat{K}_L$ and $\hat{K}_R$
respectively:
\begin{equation}\label{eq:1}
\begin{array}{lcl}
  | u_1 \rangle & = & e^{-i \frac{4 \pi
l_1}{\lambda}} \hat{K}_L | v_2 \rangle, \\
  | u_2 \rangle & = & e^{-i \frac{4 \pi
l_3}{\lambda}} \hat{K}_R | v_1 \rangle.
\end{array}
\end{equation}
At this point the two sub-cavities are still uncoupled.  In Eq.
(\ref{eq:1})  $\hat{K}_L = \hat{K}(l_1)$, $\hat{K}_R =
\hat{K}(l_3)$, where $l_1$ and $l_3$ are the lengths of the left
and right cavity respectively and  the coordinate representation
of the paraxial propagator is \cite{SiegmanBook}
\begin{equation}\label{eq:2}
\langle y | \hat{K}(l) |y' \rangle =  \sqrt{\frac{i}{B
\lambda}}\exp \left[ -i \frac{\pi}{B \lambda} \left( A {y'}^2 - 2
y y' + D y^2
 \right)
\right].
\end{equation}
The three coefficients $A,D,B$ are the corresponding elements of
the following $ABCD$ matrix:
\begin{equation}\label{eq:3}
\mathbf{M}(l) = \left(
\begin{array}{cc}
  1 - \frac{2 l}{R} & 2l(1-\frac{ l}{R}) \\
  \frac{2 }{R} & 1 - \frac{2 l}{R}
\end{array} \right),
\end{equation}
where $A = D$.

In order to describe the coupling between the two half cavities we
introduce the four scattering operators $\hat{S}_{ij}$ $(i,j =
1,2)$
\begin{equation}\label{eq:4}
\begin{array}{lcl}
  | v_1 \rangle & = & \hat{S}_{11} | u_1 \rangle + \hat{S}_{12} | u_2 \rangle, \\
  | v_2 \rangle & = & \hat{S}_{21} | u_1 \rangle + \hat{S}_{22} | u_2 \rangle,
\end{array}
\end{equation}
where the diagonal operators $\hat{S}_{ii}$ describe the
transmission of the field above the central mirror ($|y| > a$)
while the off-diagonal operators $\hat{S}_{ij}$ $(i \neq j)$
describe the reflection on the central mirror ($|y| < a$). We
require that the coupling between the two half cavities is unitary
by imposing:
\begin{equation}\label{eq:5}
\langle u_1 | u_1 \rangle + \langle u_2 | u_2 \rangle = \langle
v_1 | v_1 \rangle + \langle v_2 | v_2 \rangle,
\end{equation}
from which it follows that:
\begin{equation}\label{eq:6}
 \sum_{j = 1}^2 \hat{S}^\dag_{ij} \hat{S}_{jk} = \hat{1} \delta_{ik}, \qquad (i,j,k =
 1,2),
\end{equation}
where $\delta_{ik}$ is the Kroneker tensor. Since the bi-convex
optical element in the center of our cavity (see Fig. 1) is
invariant with respect to the symmetry $z \rightarrow -z$, we can
assume that the coupling is the same going from left to right and
viceversa, and put:
\begin{equation}\label{eq:7}
 \hat{S}_{11} =  \hat{S}_{22} \equiv  \hat{T}, \qquad  \hat{S}_{12} =  \hat{S}_{21} \equiv
 \hat{R},
\end{equation}
from which it follows that the unitarity conditions Eq.
(\ref{eq:6}) become:
\begin{equation}\label{eq:8}
\begin{array}{ccc}
  \hat{T}^\dag \hat{T} +    \hat{R}^\dag \hat{R} & = & 1, \\
  \hat{T}^\dag \hat{R} +    \hat{R}^\dag \hat{T}  & = & 0.
\end{array}
\end{equation}
Before investigating the consequences of these relations we
 collect the
four fields $u_1, u_2$ and $v_1, v_2$ in doublets
\begin{equation}\label{eq:8b}
\{ u_1, u_2 \} \rightarrow \left(
      \begin{array}{c}
                       | u_1 \rangle \\ | u_2 \rangle
      \end{array}
 \right), \qquad \{ v_1, v_2 \} \rightarrow
\left(
      \begin{array}{c}
                       | v_1 \rangle \\ | v_2 \rangle
      \end{array}
\right),
%
\end{equation}
which represent the incoming and outgoing fields in the plane
$z=0$ respectively. Alternatively is possible to relate the fields
in the left side of the cavity  $\{u_1, v_2\}$ with the fields on
the right side $\{v_1, u_2\}$ by introducing a set of four
transmission operators that are related in a simple way to the
scattering operators \cite{Spreeuw92a}. However, we prefer to use
the scattering formalism. Now we can rearrange the previous
Eqs.(\ref{eq:1}-\ref{eq:4}) as
\begin{equation}\label{eq:9}
\begin{array}{ccc}
 \left(
      \begin{array}{c}
                       | u_1 \rangle \\ | u_2 \rangle
      \end{array}
 \right)
 & =  &
\left(
       \begin{array}{cc}
                         0 & e^{-i \frac{4 \pi l_1}{\lambda} } \hat{K}_L \\
                         e^{-i \frac{4 \pi l_3}{\lambda} } \hat{K}_R & 0 \
       \end{array}
\right)
\left(
      \begin{array}{c}
                       | v_1 \rangle \\ | v_2 \rangle
      \end{array}
\right),
\end{array}
\end{equation}
and
\begin{equation}\label{eq:10}
\begin{array}{ccc}
 \left(
      \begin{array}{c}
                       | v_1 \rangle \\ | v_2 \rangle
      \end{array}
 \right)
 & =  &
\left(
       \begin{array}{cc}
                         \hat{T} & \hat{R} \\
                         \hat{R} & \hat{T} \\
       \end{array}
\right)
\left(
      \begin{array}{c}
                       | u_1 \rangle \\ | u_2 \rangle
      \end{array}
\right),
\end{array}
\end{equation}
respectively. Inserting Eq. (\ref{eq:10}) in Eq. (\ref{eq:9}) we
obtain, after a few straightforward algebraic manipulation, and
assuming the simpler case $l_1 = l_3 \equiv l \Rightarrow
\hat{K}_R = \hat{K}_L \equiv \hat{K}$, the eigenvalue equation for
the modes of the cavity:
\begin{equation}\label{eq:11}
\begin{array}{ccc}
\left(
       \begin{array}{cc}
                   \hat{R} \hat{K} - \gamma \hat{1}  &   \hat{T} \hat{K} \\
                   \hat{T} \hat{K}  &  \hat{R} \hat{K} -  \gamma \hat{1} \
       \end{array}
\right)
\left(
      \begin{array}{c}
                       | v_1 \rangle \\
                       | v_2 \rangle
      \end{array}
\right)& = & 0,
\end{array}
\end{equation}
where we defined the eigenvalue $ \gamma $ as: $\gamma = \exp (i
\frac{4 \pi l}{\lambda})$. By inspecting  Eq. (\ref{eq:11}) we can
easily recognize that the product $\hat{R} \hat{K} \equiv
\hat{K}_{RT}$ is the well known {\em round-trip} propagator
\cite{Berry03a} for a single sub-cavity. Moreover we notice that
when $\hat{T} = 0$ we get two independent eigenvalue equations for
the two unstable sub-cavities; in this case $\hat{K}_{RT}$ is not
longer unitary and $|\gamma| < 1$. With Eq. (\ref{eq:11}) we have
achieved the goal of this paper. This equation can either be
solved numerically by diagonalizing the matrix in Eq.
(\ref{eq:11}) or by applying asymptotic methods \cite{Berry01b}.
\\
In order to write Eq. (\ref{eq:11}) in coordinate representation
is necessary to write down the explicit form for the transmission
$\hat{T}$ and the reflection $\hat{R}$ operators. To this end we
first notice that the paraxial propagator which accounts for the
reflection by a convex mirror has the following  coordinate
representation:
\begin{equation}\label{eq:12}
\langle y | \hat{r} |y' \rangle =  \exp \left(  - \frac{2 \pi i
}{r \lambda}y^2 \right) \delta(y - y'),
\end{equation}
where $r$ is the radius of the convex mirror \cite{SiegmanBook}.
Since the reflection operator is a mathematical representation of
the bi-convex mirror whose transverse dimension is $2a$, its
coordinate representation must be limited to the region $|y|\leq
a$. Analogously it is easy to understand that the transmission
operator can  only exists in the region $|y|
> a$. These physical considerations  make it natural to try the
following expressions for the transmission and reflection
operators:
\begin{equation}\label{eq:13}
\begin{array}{lcl}
\langle y | \hat{T} |y' \rangle & = & \delta(y - y') \Theta(|y| - a), \\
\langle y | \hat{R} |y' \rangle & = & \delta(y - y') \Theta(a -
|y|) \exp \left(  - \frac{2 \pi i }{r \lambda}y^2 \right).
\end{array}
\end{equation}
It is easy to check, by straightforward calculation, that choosing
this form for the $\hat{R}$ and $\hat{T}$ operators, Eqs.
(\ref{eq:8}) are automatically satisfied because of the following
properties of the $\Theta$ functions:
\begin{equation}\label{eq:14}
\begin{array}{rcl}
  \Theta(|y| - a) + \Theta(a - |y|) & = & 1, \\
  \Theta(|y| - a) \Theta(a - |y|) & = & 0.
\end{array}
\end{equation}
In conclusion, we have derived the equations for a pair of coupled
unstable cavities. We obtained an eigenvalue equation
(\ref{eq:11}) which can be solved in straightforward way to get
the spectrum and the eigenmodes of the whole cavity. The theory in
the present form involves some not well defined quantities (as
products of distribution functions) which are justified only on a
physical basis.

This project is part of the program of  FOM and is also supported
by the EU under the IST-ATESIT contract.

\section{Appendix}

In this appendix we give some details about practical
calculations. We start rewriting Eq. (\ref{eq:11}) as
\\
\begin{equation}\label{eq:100}
      \begin{array}{rcl}
                   \gamma v_1(y) & = & \langle y |\hat{R} \hat{K} | v_1 \rangle
                    +     \langle y |\hat{T} \hat{K} | v_2
                    \rangle,
                    \\\\
                   \gamma  v_2(y) & = & \langle y |\hat{T} \hat{K} | v_1 \rangle +
                    \langle y |\hat{R} \hat{K} | v_2 \rangle.
                        \
\end{array}
\end{equation}
\\ For simplicity we define $\hat{R} \hat{K} \equiv \hat{\rho}$ and
$\hat{T} \hat{K} \equiv \hat{\tau}$ and write explicitly Eq.
(\ref{eq:100}) as:
%
%
\\
\begin{equation}\label{eq:110}
      \begin{array}{rcl}
                   \gamma v_1(y) & = & \displaystyle{\int dy' \rho(y, y') v_1
                   (y')
                    +     \int dy' \tau( y,y') v_2(y')},
                    \\\\
                   \gamma  v_2(y) & = & \displaystyle{ \int dy' \tau(y,y')v_1(y') +
                   \int dy' \rho(y,y') v_2(y')},
                        \
\end{array}
\end{equation}
\\
where we have defined $\rho(y, y') \equiv \langle y |\hat{\rho}|y'
\rangle$ and $\tau(y, y') \equiv \langle y |\hat{\tau}|y'
\rangle$. For a symmetrical cavity with $l_1 = l_3$ and we look
for a solution such that $v_1(y) = v_2(y)$, therfore Eqs.
(\ref{eq:110}) reduce to a single equation
\\
\begin{equation}\label{eq:120}
      \begin{array}{rcl}
                   \gamma v_1(y)  & = & \displaystyle{ \int dy' \left[ \rho(y, y') + \tau( y,y') \right]
                   v_1(y')},\\\\
                   & = & \displaystyle{ \left[ \Theta(a - |y|) \xi(y) + \Theta(|y| - a) \right]\int dy'
                   K(y,y')  v_1(y')},
\end{array}
\end{equation}
\\
where we have defined $ \xi(y) \equiv  \exp \left( - \frac{2 \pi i
}{r \lambda } y^2 \right)$. Here $ K(y,y')$ is the propagator from
a round-trip inside one unstable sub-cavity without accounting for
the reflection on the convex mirror. Instead the product $\xi(y)
K(y,y')\equiv K_{RT}(y,y')$ gives us the propagator for a {\em
complete} round-trip. For computational reasons is more convenient
to work with $K_{RT}(y,y')$ instead of $K(y,y')$ therefore,
exploiting the fact that $|\xi(y)|^2=1$ we rewrite Eq.
(\ref{eq:120}) as
\\
\begin{equation}\label{eq:130}
                   \gamma v_1(y)   =  \displaystyle{ \left[ \Theta(a - |y|)  + \Theta(|y| - a) \xi^*(y) \right]\int dy'
                   K_{RT}(y,y')  v_1(y')}.
\end{equation}
\\
After scaling all lengths with $a$, Eq. (\ref{eq:130})can be
written as
\\
\begin{equation}\label{eq:250}
                   \gamma g(y)   =  \displaystyle{\sqrt{\frac{i t}{\pi}} \left[ \Theta(1 - |y|)  + \Theta(|y| - 1) \xi^*(y) \right]
                    \int_{- \infty}^{\infty}
                   e^{- i t \left(x - y/M  \right)^2} g(x)} dx,
\end{equation}
\\
where, following Horwitz \cite{Horwitz73a}, we have defined:
\\
\begin{equation}\label{eq:260}
\begin{array}{rcl}
  M & = & \frac{\left[ \sqrt{(l+r)(R - l)} + \sqrt{l(R - r - l)} \right]^2}{r R}, \\
  F & = & \frac{a^2}{2 l \lambda (1 - l/R)}, \\
  t & = & \pi M F, \\
  \gamma & = & \gamma M^{-1/2}, \\
  g(y) & = & e^{i \pi F(M - M^{-1})/2 y^2} v(y).
\end{array}
\end{equation}
\\
The magnification $M$ can be also written in term of $m =
(A+D)/2$, the half of the trace of the $ABCD$ matrix, as $M = m +
\sqrt{m^2 -1}$.
  In
practice we have to calculate the asymptotic form of the following
three integrals:
\\
\begin{equation}\label{eq:270}
      \begin{array}{rcl}
         I_1 & = &  \displaystyle{ \int_{1}^{\infty}  e^{- i t \left(x - y/M  \right)^2} g(x) dx},\\\\
         I_2 & = &  \displaystyle{ \int_{-1}^{1}  e^{- i t \left(x - y/M  \right)^2} g(x) dx } ,\\\\
         I_3  & = & \displaystyle{ \int_{-\infty}^{-1} e^{- i t \left(x - y/M  \right)^2} g(x) dx }.
\end{array}
\end{equation}
The value $x = y/M$ (with $M>1$) for which the phase is stationary
can be inside or outside the domain of integration depending on
the value of $y$ as illustrated in the following table:

\begin{table}
\caption{\label{ta_1} The real axis $(-\infty < y < \infty)$ has
been divided in five subsets. For each of them the letters Y/N
indicate if the stationary point is contained/not contained within
the domain of integration of the integrals $I_1, I_2$ and $I_3$.}
\bigskip
\begin{tabular}{|c|c|c|c|c|c|}
  \hline
      & $-\infty<y< -M$   &  $-M<y< -1$ & $-1<y<1$ & $1<y< M$ & $M<y< \infty$ \\
     \hline \hline
  $I_1$ & N & N & N & N & Y \\  \hline
  $I_2$ & N & Y & Y & Y & N \\  \hline
  $I_3$ & Y & N & N & N & N \\
  \hline
\end{tabular}
\end{table}
%

\bibliography{CT_3}

\newpage

\begin{figure}[!h]
\includegraphics[angle=0,width=7.5truecm]{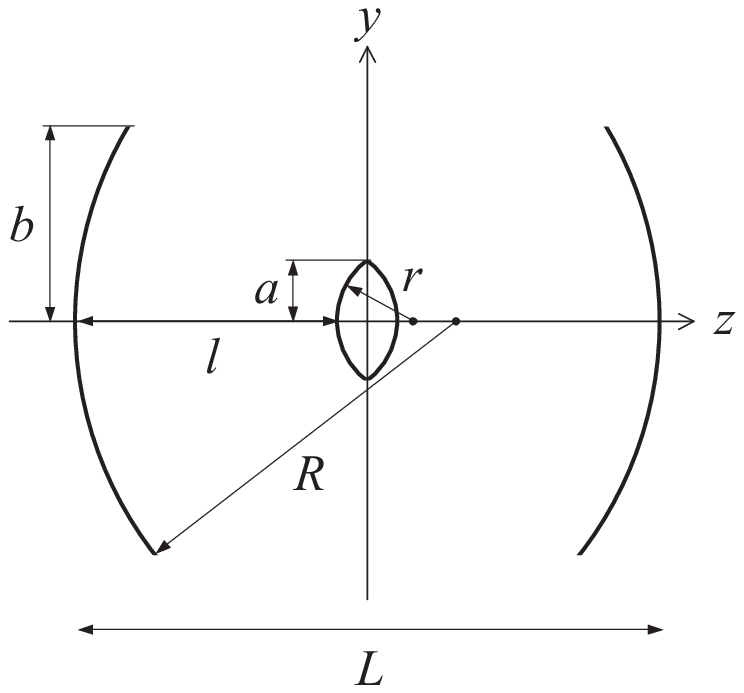}
\caption{\label{fig:1} Schematic diagram of the cavity model. Two
unstable cavities are coupled to form a single cavity which is
globally stable for $L < 2 R$. The two sub-cavities are unstable
for $l < R -r$ and stable for $R-r<l<R$.}
\end{figure}

\begin{figure}[!h]
\includegraphics[angle=0,width=7.5truecm]{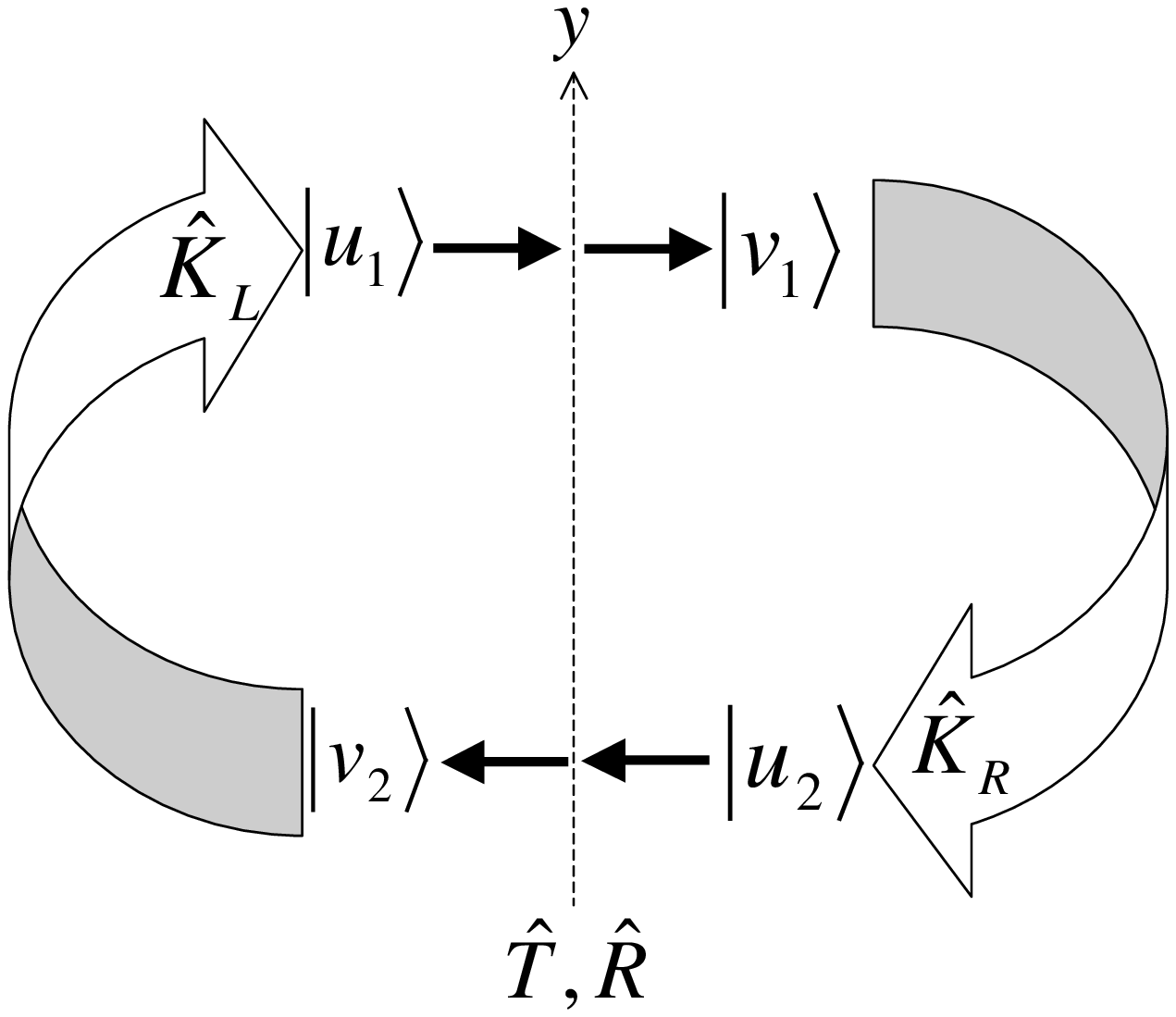}
\caption{\label{fig:2} Logical scheme of the propagation process
and of the coupling between the two sub-cavities. The dashed line
represent the plane $z = 0$ where the bi-convex mirror is located.
$\hat{K}_L$, $\hat{K}_R$ are the operators describing the field
propagation in the left and right side of the whole cavity while
$\hat{T}$ and $\hat{R}$ describe the coupling between the two
sub-cavities.}
\end{figure}

\end{document}